\documentclass{article}
\usepackage[dvips]{graphicx,color}
\usepackage{amsmath,amssymb}   
\newcommand{\bce}{\begin{center}}
\newcommand{\ece}{\end{center}}
\newcommand{\be}{\begin{equation}}
\newcommand{\ee}{\end{equation}}
\newcommand{\bea}{\begin{eqnarray}}
\newcommand{\eea}{\end{eqnarray}}
\newcommand{\bdes}{\begin{description}}
\newcommand{\edes}{\end{description}}
\newcommand{\bit}{\begin{itemize}}
\newcommand{\eit}{\end{itemize}}
\newcommand{\btt}{\begin{tt}}
\newcommand{\ett}{\end{tt}}

\def\E{\> = \>}
\def\EA{&=&}
\def\non{\nonumber\\}

\newcommand{\Eqi}{\> \equiv \>}
\def\fb{{\bf b}}  

\def\fk{{\bf k}}
\def\fK{{\bf K}}
\def\fp{{\bf p}}
\def\fq{{\bf q}}

\def\fv{{\bf v}}

\def\fw{{\bf w}}
\def\fx{{\bf x}}

\def\vecrho{\mbox{\boldmath$\rho$}}

\def\zerointT{\int_0^T dt}
\def\Iint{\int_{-\infty}^{+\infty} dt}
\def\Def{\> := \>}
\def\deF{\> =: \>}
\def\la{\left\langle \,}
\def\ra{\, \right\rangle}
\def\lrp{\left ( \, }    
\def\rrp{\, \right ) }   
\def\lsp{\left [ \, }    
\def\rsp{\, \right ] }   
\def\lcp{\left \{ \, }   
\def\rcp{\, \right \} }  
\def\A{\mathbb A}
\def\B{\mathbb B}

\begin{document}
\begin{flushright}
PSI-PR-11-02
\vspace{0.1cm}

Phys. Lett. {\bf A 375} (2011) 3781
\end{flushright}

\thispagestyle{empty}

\setcounter{page}{0}

\vspace{2cm}

\bce
{\Large\bf A New Path-Integral Representation of the $T$-Matrix in Potential Scattering} \\
\vspace*{0.3cm}

\vspace{1cm}

{\large J.~Carron$^1$ and R.~Rosenfelder$^2$}\\

\ece
\vspace{1cm}

\noindent
$^1$ Institute for Astronomy, ETH Zurich, Wolfgang-Pauli-Strasse 27, 
CH-8093 Z\"urich, Switzerland
\vspace{0.1cm}

\noindent
$^2$ Particle Theory Group, Paul Scherrer Institute,
CH-5232 Villigen PSI, Switzerland\\

\vspace{2cm}

\begin{abstract}
\noindent
We employ the method used by Barbashov and collaborators in Quantum Field Theory to derive a 
path-integral representation of the $T$-matrix in nonrelativistic potential scattering 
which is free of functional integration over fictitious variables as was necessary before. 
The resulting expression serves as a starting point for a variational approximation 
applied to high-energy scattering from a Gaussian potential. 
Good agreement with exact partial-wave calculations is found even 
at large scattering angles. A novel path-integral representation of the scattering length 
is obtained in the low-energy limit.

\end{abstract}
\newpage


{\bf 1.} 
Path integral methods are not only useful in bound-state problems but also give new insights and 
approximation schemes in quantum scattering. Recently two path-integral representations for the 
$T$-matrix in nonrelativistic potential scattering have been derived in Ref. \cite{Rose} starting
from the $S$-matrix as (infinite) limit of the time evolution operator in the interaction picture.
For a nonrelativistic particle of mass $m$ moving in a potential $V(\fx) $ they take the form
\be
{\cal T}_{i \to f} \E i \, \frac{K}{m}
\> \int d^2 b \> e^{- i \fq \cdot \fb } \> \Bigl [ \, S(\fb) - 1 \, \Bigr ] 
\label{T level 0}
\ee
where 
\be
\fK \E \frac{1}{2} \left ( \fk_i + \fk_f \right ) \> \> , \> \> K \equiv |\fK| = k 
\cos \frac{\theta}{2} \> , \qquad
\fq \E \fk_f - \fk_i \> \> , \> \> q \equiv |\fq| = 2 k \sin \frac{\theta}{2}
\ee
are the mean momentum  and momentum transfer, respectively. $ E = k^2/(2m) $ is the scattering energy 
(we set $\hbar = 1$) and $\theta$ the scattering angle. 

The main features of these representations are functional integrations  over velocities without 
boundary conditions and the use of ``phantom'' 
degrees of freedom to get rid of explicit phases which would diverge in the limit of large scattering 
times. Two versions exist which are distinguished by the reference path along which the particle 
dominantly travels and the dimensionality $d$ of the ``anti-velocity'' $\fw(t)$ which is needed 
to achieve the cancellation \footnote{Here and in the following the path 
integrals are normalized such that the free (Gaussian) integral is unity. Our notation indicates that 
$\chi$ is a function of the impact parameter $\fb$ but a 
functional of $\fv(t)$ and $\fw(t)$. Similarly for $\fx_{\rm quant}$.}.
\bea
S (\fb) \EA \int {\cal D}^3 v \, {\cal D}^d w \> \exp \left \{ \,
i \int_{-\infty}^{+\infty} dt \>  \frac{m}{2} \left [ \fv^2(t) -\fw^2(t) \right ] \, 
\right \} \>  e^{ i \, \chi(\fb,\fv,\fw]} \\
\chi(\fb,\fv,\fw] \EA \int_{-\infty}^{+\infty} dt \> V \biggl ( \, \fb + \fx_{\rm ref}(t) + 
\fx_{\rm quant}(t,\fv,\fw] \, \biggr )
\label{PI for S(b)}
\eea
In the first case 
\be
\fx_{\rm ref}^{(d=3)} (t) \E \frac{\fK}{m} \, t \> , \qquad \fx_{\rm quant}^{(d=3)}(t,\fv,\fw] 
\E \fx_{\fv}(t) - \fx_{\fw}(0)
\ee
the reference path is a straight-line path along the mean momentum and
\be
\fx_v(t) \E \frac{1}{2} \int_{-\infty}^{+\infty} dt' \> {\rm sgn}(t-t') \, \fv(t') \> , \> \> 
\dot \fx_v(t) \E \fv(t) 
\ee
describes the quantum fluctuations, $ \> {\rm sgn}(x) = 2 \Theta(x) - 1 \> $ is the sign-function.
In the second case the anti-velocity is only 1-dimensional 
\be
\fx_{\rm ref}^{(d=1)} (t) \Eqi \fx_{\rm ray}(t)\E \left [ \, \frac{\fk_i}{m} \, \Theta(-t) +  
\frac{\fk_f}{m} \, \Theta(t) \, 
\right ] \, t \> , \qquad \fx_{\rm quant}^{(d=1)}(t,\fv,w] \E \fx_{\fv}(t) - 
\fx_{\perp \, \fv}(0) - \hat \fK \, x_{\parallel \, w}(0) 
\label{ray}
\ee
and the reference path is a {\it ray} along 
the initial momentum for $ t < 0 $ and along the final momentum for $t > 0 $.
In Ref. \cite{Rose} it has been shown 
that both these representations reproduce the Born series to all orders if expanded in powers 
of the potential.
\vspace{0.3cm}

Is it possible to get rid of the ``anti-velocity'' altogether? In the present Letter we show that
this is feasible by using a method which has been introduced by Barbashov and coworkers 
\cite{BKMS,BKMPST,MaTa} nearly 
40 years ago. They showed how one can amputate exact Green functions in Quantum Field Theory 
without using perturbation theory but their work doesn't seem to have received much attention 
apart from the (rather specialized) field 
of quantum gravity where Fabbrichesi, Han and others 
\cite{FPVV,HaPo,Han,HaXu} have extensively used the Barbashov method to describe gravitational 
scattering. Our main aim in this work is to transcribe 
their method to the nonrelativistic case
and to use the result for an approximate variational calculation similar to the one investigated in 
Refs. \cite{Carr,CaRo}. As a by-product a new path-integral representation of the scattering length 
is obtained.

\vspace{0.8cm}


{\bf 2.} Our starting point is the following ``reduction formula'' for obtaining the on-shell 
$T$-matrix from
the full Green function
\be
\la \fk_f\,  \left | \, \hat T \,\right  | \, \fk_i \ra \Bigr |_{\fk_i^2=\fk_f^2 = 2m E} \> \E \> 
\lim_{\fk_i^2, \fk_f^2 \to 2 m E} \> \la \fk_f \, \left  | \, \lrp E-\hat H_0 \rrp \, \hat G(E) \, 
\lrp E-\hat H_0 \rrp\, \right  | \, \fk_i \ra \> .
\label{nonrel LSZ}
\ee
This is easily proved by using the definition of the exact Green function
\be
\hat G(E) \Def  \frac{1}{E -\hat H + i0} \E \frac{1}{E -\hat H_0 - \hat V + i0} 
\label{full G(E)}
\ee
and some standard algebraic manipulations:
\bea
\lrp E - \hat H_0 \rrp \, \hat G(E) \,\lrp E - \hat H_0 \rrp \EA \lrp E - \hat H + \hat V \rrp 
 \, \hat G(E) \,\lrp E - \hat H_0 \rrp \E \lrp 1 + \hat V \, \hat G(E) \rrp \,  
\lrp E - \hat H + \hat V \rrp \non
\EA E - \hat H_0 + \hat V \, \lrp  1 + \hat G(E) \, \hat V \rrp \Eqi  E - \hat H_0 + \hat T \> .
\label{deriv LSZ}
\eea
When going to the energy shell the first term on the r.h.s. of Eq. \eqref{deriv LSZ} vanishes and
we obtain indeed the matrix element of the $T$-matrix between initial and 
final momentum states.
Eq. \eqref{nonrel LSZ} is the nonrelativistic counterpart of the standard 
Lehmann-Symanzik-Zimmermann (LSZ) reduction formula given in any textbook on Quantum Field Theory
(see, e.g. Eq. (7.42) in Ref. \cite{PeSch}). As other tools are available, 
this relation seems not to be
employed in standard nonrelativistic scattering theory. One exception is Ref. \cite{BjDr} where
the time-dependent version in Eq. (16.82)  is used to illustrate the ``close analogy'' 
of  the reduction formula ``with the $S$ matrix constructed in propagator theory''.

\vspace{0.4cm}
Assuming that the interaction $\hat V$ is given by a local potential $ V(\hat \fx) $ we
now employ the Schwinger representation of the full Green function \eqref{full G(E)}
\be
\hat G(E) \E - i \, \int_0^{\infty} dT \>  e^{i \lrp E - \hat H + i0 \rrp \, T } 
\ee
and the velocity path integral for the time evolution operator as given in Eq. (4) of 
Ref. \cite{Rose}.
Then one obtains from Eq. \eqref{nonrel LSZ} \footnote{Our scattering states are 
normalized to $ \> \la f \,  | \, i \ra 
= (2 \pi)^3 \, \delta^{(3)} \lrp \fk_f - \fk_i \rrp $.}
\bea 
{\cal T}_{i \to f} \EA (-i) \lim_{\fk_i^2,\fk_f^2 \to 2mE} \lrp E - \frac{\fk_i^2}{2m} \rrp \, \lrp E - 
 \frac{\fk_f^2}{2m}  \rrp \, 
\,  \int_0^{\infty} dT \> e^{ i E T}
\int d^3 x \, d^3 x' \> e^{-i \fk_f \cdot \fx'+ i  \fk_i \cdot \fx } \non
&& \hspace{-1cm} \times 
\int {\cal D}^3 v\>              
\delta^{(3)} \left ( \, \fx' - \fx -\int_0^T d\tau \> 
\fv(\tau)\, \right ) \> 
\exp \lcp \, i \int\limits_0^T dt \> \lsp  \frac{m}{2} 
\fv^2(t) -   V \lrp \fx +  \int_0^{t} d\tau \> 
\fv(\tau)\,  \rrp \rsp \> \rcp  \>  . 
\eea
Performing the $\fx'$-integration
and shifting the velocity $ \> \fv(\tau) \E \bar \fv(\tau) + \fk_f/m \> $
yields
\bea
{\cal T}_{i \to f} \EA (- i) \, \lim_{\Delta_i, \Delta_f \to 0} \, \lrp \Delta_i \,  
\Delta_f \rrp \, \cdot \, 
 \int_0^{\infty} dT \> e^{i \Delta_f T} 
\, \int d^3 x \> e^{-i (\fk_f -\fk_i) \cdot \fx} \non
&&  \times \int {\cal D}^3 \bar v \>             
\exp \lcp \, i \int\limits_0^T dt \> \lsp  \frac{m}{2} 
\bar \fv^2(t) -   V \lrp \fx + \frac{\fk_f}{m} t + \int\limits_0^{t} d\tau \> 
\bar \fv(\tau)\,  \rrp \rsp \> \rcp  
\label{T 1}
\eea
where
\be
\Delta_i \Def E - \frac{\fk_i^2}{2m} \> , \qquad \Delta_f \Def E - \frac{\fk_f^2}{2m} 
\label{def Delta}
\ee
is the initial or final off-shellness.
It is obvious that Eq. \eqref{T 1} is only non-zero if the Fourier-transformed Green function 
develops {\it two} poles when $\fk_i^2, \fk_f^2 \to 2m E $. How is that possible? 
It only can originate from the integral
over $T$, presumably from $T \to \infty$ (one intuitively expects that the particle becomes 
on-(energy-)shell 
for infinite times). Indeed for any smooth function $f(T)$ one has
\bea
-i \,  \lim_{\Delta \to 0 } \,   \Delta \, \int_0^{\infty} dT \> 
e^{i \, ( \Delta  +i0 ) \, T } \, f(T) \EA
- \lim_{\Delta \to 0} \, \int_0^{\infty} dT \> \frac{\partial}{\partial T}\, 
\lcp e^{i \, ( \Delta +i0 ) \, T } \rcp \, f(T) \non
&\stackrel{\rm (partial \, int.)}{=} &  f(0) \, + \, \lim_{\Delta \to 0} \, 
\int_0^{\infty} dT \>  
e^{i \, \Delta  T } \>
\frac{\partial f(T)}{\partial T} \E f(\infty) \> .
\label{trunc}
\eea
However, integration over $T$ only takes away the factor $ \Delta_f $ but 
{\it not} the other
factor  $ \Delta_i $  and one obviously needs a second integration.
The main trick in Barbashov's treatment
is a way to provide for this second time integration by first subtracting from  
Eq. \eqref{T 1} 
a term (a ``1'') which does not contribute in the on-shell limit 
\bea
{\cal T}_{i \to f} \EA  - i \, \lim_{\Delta_i, \Delta_f \to 0 } \, \Delta_i \, \Delta_f \,  
\int_0^{\infty} dT \> e^{ i \lrp \Delta_f  + i0 \rrp \, T } \int d^3 x \> e^{-i \fq \cdot \fx} \non
&& \times  \int {\cal D}^3 v \>  \exp \lrp i  \zerointT \> \frac{m}{2} \fv^2(t) \rrp \>          
\lcp   \exp \lsp  - i \zerointT \> V ( \vecrho_f(\fx,t,\fv] ) \rsp
\, - 1 \rcp \> .
\label{T 2}
\eea
This is because 
the free Green function is proportional to $ \> \delta^{(3)} (\fk_f-\fk_i)/(E - \fk_i^2/(2m)) \> $. 
We use the abbreviation
\be
\vecrho_f(\fx,t,\fv] \Def  \fx + \frac{\fk_f}{m} \, t \, + \, \int_0^t d\tau \> \fv(\tau) 
\ee
and omit the bar on the velocity variable.
Now the factor in curly brackets in Eq. \eqref{T 2} can be written as
\be
\exp \lsp  - i \, \int_0^T dt \> V (\vecrho_f(\fx,t,\fv] ) \rsp
\> - 1  \E  \lsp  - i \,  \int_0^T d\xi \> V ( \vecrho_f(\fx,\xi,\fv] ) \rsp \, 
\int_0^1 d\lambda \>  \exp \lsp  -i \lambda \,  \int_0^T dt \> V (\vecrho_f(\fx,t,\fv] ) \rsp
\ee
which provides an additional integration over the time $\xi$.
\vspace{0.6cm}


\noindent
{\bf 3.} Since the algebra in Ref. \cite{HaPo} (which we follow with several corrections) is
somewhat sketchy and involved we list in more detail the steps to be taken in the following:

\bdes
\item[(i)] Write
\be
\int_0^{\infty} dT \> \int_0^T d\xi \> ... \E \int_0^{\infty} d\xi \> \int_{\xi}^{\infty} dT \> ... 
\ee
and substitute $  \> T \deF \xi + s  \> $ so that $ s \in [ 0,\infty ] \>.$
We then have
\bea
{\cal T}_{i \to f} \EA (-i)^2 \,
\lim_{\Delta_i, \Delta_f \to 0} \Delta_i \, \Delta_f \, 
\int_0^{\infty} d\xi \,  \int_0^{\infty} ds \> 
e^{i \lrp  \Delta_f +i0 \rrp \, (\xi + s)} \, \int {\cal D}^3 v \,
 \exp \lrp \, i \int\limits_0^{\xi+s} dt \,  \frac{m}{2} \fv^2(t) \rrp  \non
&& \times \int d^3 x \>  e^{- i \fq\cdot \fx} \, V \lrp \vecrho_f(\fx,\xi,\fv] \rrp \, 
\, \int_0^1 d\lambda \> \exp \lsp  - i \lambda  \int_0^{\xi+s} dt \> 
V \lrp  \vecrho_f(\fx,t,\fv] \rrp \rsp \> .
\label{T 3}
\eea

\item[ii)]  Substitute
\be
\bar \fx  \E \vecrho_f(\fx,\xi,\fv] \E \fx +\frac{\fk_f}{m} \, \xi +  \int_0^{\xi} d\tau \> 
\fv(\tau) \>.
\ee
Collecting the pieces in the exponents we then obtain
\bea
{\cal T}_{i \to f} \EA  (-i)^2   \,
\lim_{\Delta_i, \Delta_f \to 0 }  \Delta_i \, \Delta_f\, 
\int_0^{\infty} d\xi \,  \exp \lsp i \lrp \Delta_i  +i0 \rrp  \xi  
+ \underline{i \frac{\fq^2}{2 m} \,  \xi} \, \rsp \, \non
&& \times 
\int_0^{\infty} ds  \, \exp \lsp  i \lrp \Delta_f + i0 \rrp  s \, \rsp
\, \int {\cal D}^3 v \,
\exp \lsp \,  i\int\limits_0^{\xi+s} d\tau \,  \frac{m}{2} \fv^2(\tau) + \underline{i \, \fq \cdot 
\int\limits_0^{\xi} d\tau \, \fv(\tau) }\, \rsp \non 
&& \times \, \int d^3 \bar \fx \> e^{-i \fq\cdot \bar \fx} \,
\> V(\bar \fx)\, 
\, \int_0^1 d\lambda \> \exp \lsp  - i \lambda\int_0^{\xi+s} dt \> 
V \lrp \bar \fx + \vecrho_f(\bar \fx,t,\fv] - \vecrho_f(\bar \fx,\xi,\fv] \rrp
\> \rsp 
\label{T 4}
\eea

\item[(iii)] The unwanted pieces (underlined in Eq. \eqref{T 4}) are 
eliminated by yet another shift
\be
\bar \fv(\tau) \E \fv(\tau) + \frac{\fq}{m} \, \Theta( \xi - \tau ) \> .
\ee
The argument in the last potential term then becomes
\be
\bar \fx + \vecrho_f(\bar \fx,t,\fv] - \vecrho_f(\bar \fx,\xi,\fv] \E
\bar \fx + \lsp \,  \frac{\fk_f}{m} \,  \Theta( t - \xi )
+  \frac{\fk_i}{m} \,  \Theta( \xi - t ) \, \rsp \, ( t - \xi )  + 
\int_{\xi}^t d\tau \> \bar \fv(\tau)\> .
\label{arg}
\ee
Finally, we define
\be
\bar \tau \Def \tau - \xi \> , \quad \bar t \Def t - \xi \> , \quad
\overline{\overline{\fv}}(\bar \tau) \Def \bar \fv(\bar \tau + \xi)
\ee
so that the argument \eqref{arg} becomes
\be
\bar \fx  + \lsp \,  \frac{\fk_f}{m}  \, \Theta( \bar t )
+  \frac{\fk_i}{m} \, \Theta( - \bar t ) \, \rsp \, \bar t + 
\int_0^{\bar t} d\bar \tau \, \overline{\overline{\fv}}(\bar \tau) \, 
 \deF  
\vecrho_{\rm ray}(\bar \fx,\bar t, \overline{\overline{\fv}}] \> .
\ee
Omitting the bars over the transformed variables we thus have
\bea
{\cal T}_{i \to f} \EA   (-i)^2 \,
\lim_{\Delta_i, \Delta_f \to 0}  \Delta_i \, \Delta_f\,
\int\limits_0^{\infty} d\xi \,  e^{i \lrp \Delta_i  +i0 \rrp  \xi } \, 
\int_0^{\infty} ds  \, e^{i \lrp \Delta_f + i 0 \rrp s }
\int d^3 x \> e^{-i \fq \cdot \fx} \, V(\fx) \non
&& \times 
\, \int {\cal D}^3 v \,
\exp \lsp \, i \int\limits_{-\xi}^s d\tau \> \frac{m}{2} \fv^2(\tau) \, \rsp 
\, \int_0^1 d\lambda \> \exp \lsp  - i \lambda   \int\limits_{-\xi}^s dt \> 
V \lrp \vecrho_{\rm ray}(\fx,t,\fv] \rrp \> \rsp \> .
\label{T 5}
\eea

\edes
It is now possible to truncate {\it both} external legs by using Eq. \eqref{trunc} and to obtain the 
final result
\be
{\cal T}_{i \to f} \E \int d^3 x \> e^{-i \fq \cdot \fx} \, V(\fx)
\, \int {\cal D}^3 v \> 
\exp \lsp \,  i  \int\limits_{-\infty}^{+\infty} dt \> \frac{m}{2} \fv^2(t) \, \rsp 
\,  \int_0^1 d\lambda \> 
\exp \lsp  - i \lambda  \int\limits_{-\infty}^{+\infty} dt \ 
V \lrp \vecrho_{\rm ray}(\fx,t,\fv] \rrp \> \rsp \> . 
\label{T Barb}
\ee
Of course, the $\lambda$-integration can be performed trivially but for many applications 
it is better to leave it in this form (see below). 

\noindent
Note that in the second potential term the particle propagates mainly along a {\it ray} 
formed by incident and outgoing momenta exactly as in the ``ray'' representation \eqref{ray}
while the quantum fluctuations are described by the functional integration over the velocity.
Indeed 
\be
\vecrho_{\rm ray}(\fx,t,\fv) \E  \fx + \fx_{\rm ray}(t) + \fx_{\fv}(t) -  \fx_{\fv}(0) \> . 
\ee
Eq. \eqref{T Barb} also gives a precise definition
for a {\it local} ``pseudopotential'' (in general, this is a nonlocal object; see, e.g. 
chapter 6.2 in Ref. \cite{RodThal}) whose Fourier transform yields the exact $T$-matrix.

While it is not possible  to set $ k =  0 $ directly in Eq. \eqref{T level 0} 
 the representation
\eqref{T Barb} allows that without impunity. Therefore, 
defining the scattering length $a$ as the negative $(k \to 0)$-limit of the scattering amplitude 
$ f = - m {\cal T}/(2 \pi) $ (as in Eq. (X.47) of Ref. \cite{Mess}) we immediately obtain
a path-integral representation of the scattering length
\be
a \E \frac{m}{2 \pi} \, \int d^3 x \, V(\fx) \, \int {\cal D}^3 v \, 
\exp \lsp i  \int_{-\infty}^{+\infty} dt \, \frac{m}{2} \fv^2(t)  \rsp 
\, \int_0^1 d\lambda \, \exp \lsp  - i \lambda  \int_{-\infty}^{+\infty} dt \,
V \lrp \fx + \int_0^t d\tau \, \fv(\tau) \rrp \, \rsp  \> .
\label{a}
\ee
Unfortunately, we do not have a simple argument
why the scattering length should be real in potential scattering, i.e.
why the imaginary part of the r.h.s of Eq. \eqref{a} vanishes. This is one of the shortcomings
of a path-integral representation of scattering; others are that 
basic properties like unitarity (and therefore 
the optical theorem) are more evident in the traditional operator formulation of 
Quantum Mechanics.

\vspace{0.8cm}


{\bf 4.} 
We can use our result as a starting point for new approximation schemes. For example,
following Refs. \cite{Carr,CaRo} we may variationally approximate the velocity path integral in Eq.
\eqref{T Barb}
\be
F(\fx,\lambda,S] \Def \int {\cal D}^3 v \>  e^{i S(\fx,\lambda,\fv]} \> , \qquad S(\fx,\lambda,\fv] \E 
\Iint \> \lsp \frac{m}{2} \fv^2(t) - \lambda 
V ( \rho_{\rm ray}(\fx,t,\fv] ) \> \rsp 
\label{F}
\ee
by means of the Feynman-Jensen variational principle. The most general quadratic trial action,
which allows the analytic evaluation of the needed path integrals, is
\be
S_t[\fv] \E \frac{m}{2} \, \fv^T \, \A \,  \fv + \B \, \fv
\ee
in the condensed notation of Ref. \cite{CaRo} 
where summation over discrete and continuous indices is implied. Evaluation of averages and 
derivation of the variational equations for the functions
$A_{ij}(t,t') , B_i(t) $ follow the same lines as in Refs. \cite{Carr,CaRo} where more details 
can be found. Here we just give the results: with this trial action
the variational approximation to the $T$-matrix with no anti-velocity reads
\be
{\cal T}^{(AB30)}_{i \to f} \E \int d^3 x \> e^{-i \fq \cdot \fx} \, V(\fx) \, \int_0^1 d\lambda \> 
F_{\rm var}(\fx,\lambda) \> , 
\qquad F_{\rm var}(\fx,\lambda)  \E  e^{ iX_0 + iX_1 - \Omega} 
\ee
with 
\begin{subequations}
\bea
X_0 \EA - \Iint  \>  U(t)  \\
X_1 \EA \frac{1}{2} \, \nabla U \, \Sigma_0 \,  \nabla U \\
\lrp \Sigma_0\rrp_{ij}(t,t')  \EA  -\frac{\delta_{ij}}{2m} \Bigl [ \> |t-t'| - |t'| - |t| \> 
\Bigr ] \\
\Omega \EA -\frac{1}{2} {\rm Tr} \lsp \ln  \lrp 1 + \Sigma H  \rrp - \Sigma H \rsp \> .
\eea
Here $ U(t)$
is the Gaussian transform of the potential $\lambda V$, i.e.
\be
U(t) \E \int \frac{d^3 p}{(2 \pi)^3} \> \lambda \tilde V(\fp) \,  \exp \lrp i \fp \cdot 
\vecrho_{\rm var}(t)
-  \frac{i}{2} \, \fp^T \, \sigma(t) \, \fp \rrp \> , \quad {\rm with} \qquad 
\tilde V(\fp) \E \int d^3 x \> 
e^{-i \fp \cdot \fx} \,  V(\fx)
\ee
\end{subequations}
and $ H_{ij} \Def \partial_i\partial_j \, U $ its Hessian. The width of the Gaussian transform
is given by the diagonal value of the matrix $\Sigma$, i.e. $ \> \sigma_{ij}(t) \E 
\Sigma_{ij}(t,t)\> $.
Varying with respect to the variational functions $ \A, \B$ leads to 
variational equations for the trajectory and the (off-diagonal) width:
\bea
\vecrho_{\rm var}(t) \EA \fx + \fx_{\rm ray}(t) + \lrp \Sigma_0 \, \nabla U \rrp(t)
\label{var for rho} \\  
\Sigma \EA  \Sigma_0  + \Sigma_0 \, 
H \, \Sigma \> .
\label{LS for Sigma}
\eea
The first one is a classical equation of motion for the trajectory of the particle as can be seen by
differentiating Eq. \eqref{var for rho} twice with respect the the time $t$.
This gives Newton's equation  with a ``kick'' at $t = 0$
\be
m \, \ddot \vecrho_{\rm var}(t) \E \lrp \fq + \Iint' \> \nabla U(t') \rrp \, \delta(t) 
- \nabla U(t)  \> .
\ee
The second one is a Lippmann-Schwinger-type equation for the quantum-mechanical spreading of the
scattered wave. In the present numerical evaluation of these coupled nonlinear equations 
we have restricted ourselves
to the case where $ \A = 1 $. This amounts to setting the Hessian to zero and to neglect the wave
spreading.
Then the width of the Gaussian transform is fixed as $ \> \sigma_{ij}(t) \E \delta_{ij}\,  |t|/m \> $
and the quantity $\Omega$ vanishes. 

This approximation (B30, in the nomenclature of Ref. \cite{CaRo})
should reproduce the $2^{\rm nd}$ Born approximation and 
therefore be roughly as good as
the systematic approximations studied in Ref. \cite{CaRo} which included the correction 
by the second cumulant. Indeed, simple algebra shows that the second Born approximation
is contained in our variational approximation: When expanding $ {\cal T}^{(B30)}_{i \to f} $
up to second order in the potential
we can neglect the phase $X_1$ and set 
$ \exp(iX_0) = 1 + i X_0 $.
Furthermore the argument of the potential in the phase $X_0$ reduces to 
$ \fx + \fx_{\rm ref}(t) $ in this order. 
The $\lambda$-integration is then trivial and with the definition of the Gaussian 
transformed potential one can perform the 
$\fx$-integration. The result is
\begin{subequations}
\be
T^{(B30)}_{i \to f} \E T^{\rm 1st\,Born} - \frac{i}{2} \int \frac{d^3p_1 d^3p_2}{(2 \pi)^3}  \, 
\tilde V(\fp_1) \, 
\tilde V(\fp_2) \ \delta^{(3)} ( \fp_1 + \fp_2 - \fq ) 
\int_{-\infty}^{+\infty} dt \, \exp \left [ i \fp_2 \cdot \fx_{\rm ray}(t) - \frac{i}{2m}
|t|\,  \fp_2^2 \right ] +  {\cal O}(V^3) 
\ee
where the Gaussian width in the the ``B30''-approximation 
has been inserted. This  
allows the $t$-integration to be performed and gives
\bea
T^{(3-0)}_{i \to f} \EA T^{\rm 1st\,Born} -  \frac{i}{2} \int \frac{d^3p_1 d^3p_2}{(2 \pi)^3}  \, 
\tilde V(\fp_1) \, 
\tilde V(\fp_2) \ \delta^{(3)} ( \fp_1 + \fp_2 - \fq ) \non
&& \times \left [ \, \frac{i}{\fp_2 \cdot \fk_f/m - \fp_2^2/(2m) + i 0 } 
-  \frac{i}{\fp_2 \cdot \fk_i/m + \fp_2^2/(2m) - i 0 }
\right ] + {\cal O}(V^3) 
\eea
where the two terms come from the integration over positive and negative times, respectively. 
Now, putting $ \> \fp_1 \E \fp - \fk_i \> , \quad \fp_2 =  \fk_f - \fp \> $ 
in the first term and just reversed in the second, one sees that both terms are the same due to 
energy conservation  $ E = \fk_i^2/(2m) =  \fk_f^2/(2m)$ which is already imposed 
by dealing with the on-shell $T$-matrix.
This cancels the factor $1/2$ from the $\lambda$-integration so that the result is
\bea
T^{(B30)}_{i \to f} \EA T^{\rm 1st\,Born} + \int \frac{d^3p}{(2 \pi)^3}  \, \tilde V(\fk_f - \fp) \, 
\frac{1}{E - \fp^2/(2m) + i 0} \, 
\tilde V(\fp- \fk_i)  + {\cal O}(V^3)  \non
\EA  T^{\rm 1st\,Born} + \left < \fk_f \left | \, \hat V \, \frac{1}{E - \hat H_0 + i 0} \, 
\hat V \, \right | 
\fk_i \right >  + {\cal O}(V^3) \equiv  T^{\rm 1st\,Born} +  T^{\rm 2nd\,Born}  + 
{\cal O}(V^3) \> .
\eea
\end{subequations}
\vspace{0.1cm}

\begin{figure}[hbt]
\vspace*{1cm}
\bce
\includegraphics[angle=90,scale=0.6]{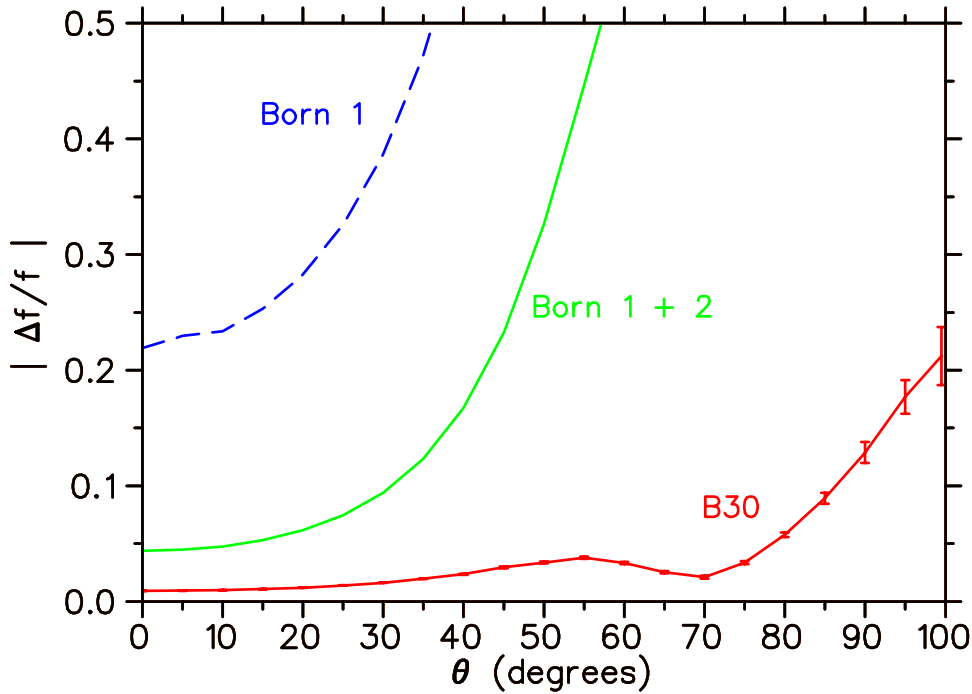}
\ece
\caption{(Color online) The relative deviation $|\Delta f/f|$ of the  
variational amplitude in the present approximation (``B30'')  from the 
exact scattering amplitude as a function of the scattering angle $\theta$. The error bars denote
the estimated uncertainty of the numerical evaluation.
Also shown are the
results from the first and from the first + second Born approximation.}
\label{fig: reldev B30}
\end{figure}

We have solved the variational equations by iteration similar as was done in Refs. 
\cite{Carr,CaRo} apart
from numerical improvements to speed up the tabulation of the trajectory and to add 
the asymptotic contribution to integrals with slowly decreasing integrands.
Fig. \ref{fig: reldev B30}  shows the comparison with an exact partial-wave calculation 
for scattering from a Gaussian potential
\be
V(x) \E V_0 \, e^{-x^2/R^2} 
\ee
at $ k R = 4 \, , \> 2m V_0 \, R^2 = - 4$ which served as a benchmark calculation in 
Refs. \cite{Carr,CaRo}. Plotted is the relative deviation $ \> | \Delta f/f | \Eqi
| \lrp f_{\rm var} - f_{\rm exact} \rrp / f_{\rm exact} | \> $ and it is seen that our variational
approximation outperforms both the first and the second Born approximation. Even at large scattering
angles where the cross section is down by many orders of magnitude the deviation is still rather small.

\vspace{1.2cm}

{\bf 5.} In summary we have presented a new path-integral representation for the $T$-matrix 
in nonrelativistic potential scattering based on an ingenious method 
for amputating full Green functions in Quantum Field Theory due to Barbashov {\it et al.}
Compared to previous representations the present one does not need
artificial ``phantom'' degrees of freedom to cancel unphysical contributions. 

Of course, there is an (infinite) variety of path-integral representations
depending at which level of formal scattering theory one introduces the path integral. For reasons 
of classification one may count the powers of the potential in front of the velocity path integral.
Then the previous forms displayed in Eqs. \eqref{T level 0}, \eqref{PI for S(b)} may be characterized 
as ``level 0''-
representations whereas the Barbashov-like result \eqref{T Barb} belongs to the ``level 1''-class.
A path-integral representation based on the formal solution of the Lippmann-Schwinger equation 
(last line in Eq. \eqref{deriv LSZ}) would lead to a ``level 2''-representation and so on. Obviously,
higher-level representations become more and more complicated so that  the formulations on level 
zero or one seem to be a good compromise between complexity and efficiency.

\vspace{0.2cm}

As shown by a variational calculation for high-energy scattering from a Gaussian potential
the new representation offers new opportunities for approximations, seems to be applicable also
at low energy and could be used for stochastic evaluation of real-time
scattering similar as in Ref. \cite{Rose2}. At least it is an interesting addition 
to the standard body of 
knowledge in nonrelativistic scattering theory originating from field-theoretic methods.



\vspace{1.5cm}

\begin{thebibliography}{99}



\bibitem{Rose} R.~Rosenfelder: 
  ``Path Integrals for Potential Scattering,''
  Phys. Rev. A {\bf 79} (2009) 012701  
    [arXiv:0806.3217 [nucl-th]].



\bibitem{BKMS} 
  B.~M.~Barbashov, S.~P.~Kuleshov, V.~A.~Matveev and A.~N.~Sisakian:
  ``Eikonal approximation in quantum-field theory'', 
    Teor. Mat. Fiz. {\bf 3} (1970) 342 
    [English transl.: Theor. Math. Phys. {\bf 3} (1970) 555] .

\bibitem{BKMPST}
  B.~M.~Barbashov, S.~P.~Kuleshov, V.~A.~Matveev, V.~N.~Pervushin, A.~N.~Sissakian 
  and A.~N.~Tavkhelidze:
  ``Straight-line paths approximation for studying high-energy elastic and
  inelastic hadron collisions in quantum field theory,''
  Phys.\ Lett.\  B {\bf 33} (1970) 484.


\bibitem{MaTa} V. A. Matveev and A. N. Tavkhelidze: ``On the representation of scattering
   amplitudes as path integrals in quantum field theory'', 
   Teor. Mat. Fiz. {\bf 9} (1971) 44
   [English transl.:  Theor. Math. Phys. {\bf 9} (1971) 968] .

\bibitem{FPVV}   M.~Fabbrichesi, R.~Pettorino, G.~Veneziano and G.~A.~Vilkovisky:
  ``Planckian energy scattering and surface terms in the gravitational
  action'':
  Nucl.\ Phys.\  B {\bf 419} (1994) 147.


\bibitem{HaPo}   N.~S.~Han and E.~Ponna: ``Straight-line path approximation for studying 
   Planckian-energy scattering in quantum gravity'', Nuov. Cim. {\bf A 110} (1997) 459.

\bibitem{Han}   N.~S.~Han:
   ``Straight-line path approximation for high-energy elastic and inelastic scattering 
     in quantum gravity''
  Eur.\ Phys.\ J.\  C {\bf 16} (2000) 547.


\bibitem{HaXu}   N.~S.~Han and N.~N.~Xuan:
  ``Planckian scattering beyond the eikonal approximation in the functional approach'',
  Eur.\ Phys.\ J.\  C {\bf 24} (2002) 643.
  [arXiv:gr-qc/0203054].


\bibitem{Carr}  J.~Carron:
  ``Variational Methods for Path Integral Scattering,''
  arXiv:0903.0273 v2 [nucl-th] .

\bibitem{CaRo}
  J.~Carron and R.~Rosenfelder:
  ``Variational Approximations in a Path-Integral Description of Potential
  Scattering,''
  Eur.\ Phys.\ J.\  A {\bf 45} (2010) 193
  [arXiv:0912.4429 [nucl-th]].



\bibitem{PeSch} M. E. Peskin and D. V. Schroeder: {\it An Introduction to Quantum Field Theory},
Addison-Wesley (Reading 1996) .

\bibitem{BjDr} J. D. Bjorken and S. D. Drell: {\it Relativistic Quantum Fields}, 
McGraw-Hill (New York 1965) .

\bibitem{RodThal} L. S. Rodberg and R. M. Thaler: {\it Introduction to the Quantum 
Theory of Scattering}, Academic Press (New York 1967) .

\bibitem{Mess} A. Messiah: {\it Quantum Mechanics}, North-Holland (Amsterdam 1965) .


 \bibitem{Rose2} 
  R.~Rosenfelder:
  ``Exact Path-Integral Representations for the $T$-Matrix in Nonrelativistic Potential Scattering,''
  Few Body Syst.\  {\bf 49 } (2011)  41
  [arXiv:1008.1718 [nucl-th]] .


\end{thebibliography}
\end{document}